\documentclass[aps,prl,print,12 pt, onecolumn, footinbib,superscriptaddress]{revtex4-2}

\usepackage{graphicx} 
\usepackage{gensymb} 

\usepackage{microtype}
\usepackage{amssymb,amsmath}
\usepackage{graphicx}
\usepackage{color}
\usepackage{listings}
\usepackage{float}
\usepackage{wrapfig}
\usepackage{amsmath}
\usepackage{gensymb}
\usepackage{soul}
\usepackage[version=4]{mhchem}
\usepackage{chemmacros}

\usepackage{upgreek}
\usepackage{enumerate} 
\usepackage[linkcolor = blue, citecolor = blue, urlcolor = blue, colorlinks = true]{hyperref}

\usepackage{commath,amssymb}
\usepackage{ushort}
\usepackage{multirow}
\usepackage{cleveref}
\usepackage{epstopdf}

\crefname{equation}{Eq.}{Eqs.}
\crefname{figure}{Fig.}{Figs.}

\usepackage[separate-uncertainty]{siunitx}

\usepackage{hyphenat}
\usepackage{booktabs}
\renewcommand{\selectlanguage}[1]{}

\usepackage{tabularx}
\usepackage{array}
\usepackage{makecell}
\setcitestyle{super,open={},close={}}

\usepackage{xr}
\makeatletter

\newcommand*{\addFileDependency}[1]{
\typeout{(#1)}
\@addtofilelist{#1}
\IfFileExists{#1}{}{\typeout{No file #1.}}
}\makeatother

\newcommand*{\myexternaldocument}[1]{%
\externaldocument{#1}%
\addFileDependency{#1.tex}%
\addFileDependency{#1.aux}%
}

\myexternaldocument{si}

\newcommand{\affleidenexp}{\affiliation{\small Huygens-Kamerlingh Onnes Laboratory, Leiden University, P.O. Box 9504, 2300 RA Leiden, The Netherlands}}
\newcommand{\affamolf}{\affiliation{\small AMOLF, 1098 XG Amsterdam, The Netherlands}}
\newcommand{\affleidensim}{\affiliation{\small Lorentz Institute, Leiden University, P.O. Box 9506, 2300 RA Leiden, The Netherlands}}

\mathcode`\,="213B 

\begin{document}
\newpage

\title{\Large\bf Colloidal Pivots Enable Brownian Metamaterials and Machines}

\author{Julio Melio}
\affleidenexp
\author{Martin van Hecke}
\affleidenexp
\affamolf
\author{Silke E. Henkes}
\affleidensim
\author{Daniela J. Kraft}
\email[Corresponding email: ]{kraft@physics.leidenuniv.nl}
\affleidenexp

\maketitle

\textbf{Biological machines harness targeted deformations that can be actuated by Brownian fluctuations. However, while synthetic micromachines can similarly leverage targeted deformations they are too stiff to be driven by thermal fluctuations and thus require
strong forcing~\cite{hu_magnetic_2021, smart_magnetically_2024,liu_electronically_2025}. Furthermore, systems that are able to change their conformation by thermal fluctuations do so uncontrollably~\cite{mcmullen_self-assembly_2022,melio_soft_2024} or require external control~\cite{aubret_metamachines_2021}. Here we leverage DNA-based sliding contacts~\cite{van_der_meulen_solid_2013, chakraborty_colloidal_2017,rinaldin_colloid_2019} to
create colloidal pivots, rigid anisotropic objects that freely fluctuate around their pivot point, and use a hierarchical strategy to assemble these into Brownian metamaterials and machines with targeted deformation modes.
We realize the archetypical rotating diamond and rotating triangle, or Kagome, geometries, and quantitatively show how thermal fluctuations drive their predicted auxetic deformations~\cite{Resch1963, Grima2000, mullin_pattern_2007, Lubensky2015, bertoldi_flexible_2017, kadic_3d_2019}. Finally, we implement magnetic particles into the colloidal pivots to achieve an elementary Brownian machine with easily actuatable deformations that can harness Brownian fluctuations. 
Together, our work introduces a strategy for creating thermal mechanical metamaterials and leverages them for functional Brownian devices, paving the way to materialize flexible, actuatable structures for micro-robots, smart materials, and nano-medicine.}

\begin{figure}[hb!]
    \centering
    \includegraphics[width=0.5\textwidth]{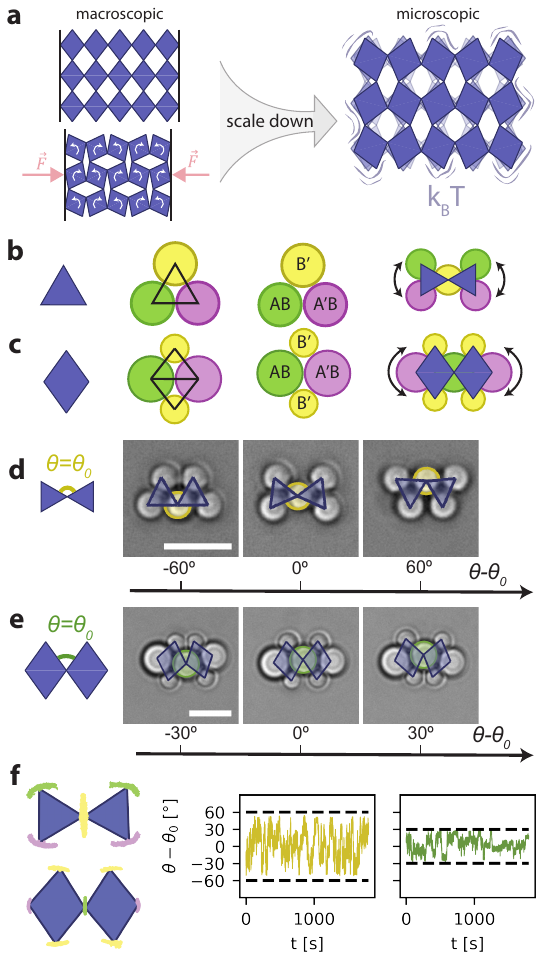}
    \caption{\textbf{The building blocks of thermal mechanical metamaterials} a) Translating mechanical metamaterials such as the rotating diamond lattice to the microscopic length scale where thermal fluctuations instead of external forces deform the structure requires b,c) creating rigid units connected by hinging bonds. We experimentally realize (b) hinging triangle and (c) diamond units by placing colloidal joints at each corner of the diamond. Colloidal spheres of two sizes and two complementary pairs of surface-mobile DNA linkers (indicated by A-A' and B-B') are used to ensure mechanical rigidity of the units while allowing hinging. (d,e) Pairs of hinging (d) triangle and (e) diamond units show random fluctuations in their relative opening angle as observed by brightfield microscopy. Scale bars are 5 $\mathrm{\mu m}$. f) Using quantitative image analysis we extract the positions of all particles in time relative to the center of mass (left) and the random changes in the opening angle which fluctuate in the geometrically allowed range (dashed lines). }
    \label{fig:fig1}
\end{figure}

Mechanical metamaterials and machines require a precise arrangement of rigid and flexible elements, such as hinges, that translate local rotations to controlled, collective deformations~
\cite{Resch1963, Grima2000, mullin_pattern_2007, Lubensky2015, bertoldi_flexible_2017,kadic_3d_2019, guo_non-orientable_2023}. 
Their miniaturization, using proteins~\cite{suzuki_self-assembly_2016}, DNA~\cite{li_auxetic_2021, li_ultrastrong_2023}, self-assembled colloidal particles~\cite{Chen2011,Mao2013,Wang2024}, replica molding~\cite{zhang_one-step_2008}, advanced 3D printing~\cite{Buckmann2012,Frenzel2017,Xia2019, zhang_hydrogel_2023,Dudek2023} or lithography~\cite{Dorsey2019,smart_magnetically_2024,liu_electronically_2025},
opens up exciting possibilities for creating functional materials~\cite{kadic_3d_2019, barrat_soft_2023, dudek_shape_2025} and microscopic robots and machines~\cite{bhattacharya_applied_2005, liu_colloidal_2023,  bishop_active_2023,smart_magnetically_2024,liu_electronically_2025}. However, in contrast to biological machines, the synthetic hinges used in these metamaterials are too stiff to respond to thermal fluctuations and require strong external actuation for their function. 
Colloidal joints--—particles with sliding contacts enabled by surface-mobile DNA linkers~\cite{chakraborty_colloidal_2017, mcmullen_self-assembly_2022, melio_soft_2024}—--can provide highly flexible bonds. However, they are not yet suited for Brownian metamaterials, which require freely-hinging bonds between rigid, non-spherical elements (Fig.~\ref{fig:fig1}a). This is because spherical colloidal joints are flexible but lead to uncontrolled deformations~\cite{melio_soft_2024} and colloidal joints between non-spherical particles lose their flexibility~\cite{chakraborty_colloidal_2017, shelke_flexible_2023}.

\begin{figure*}
    \centering
    \includegraphics[width=\textwidth]{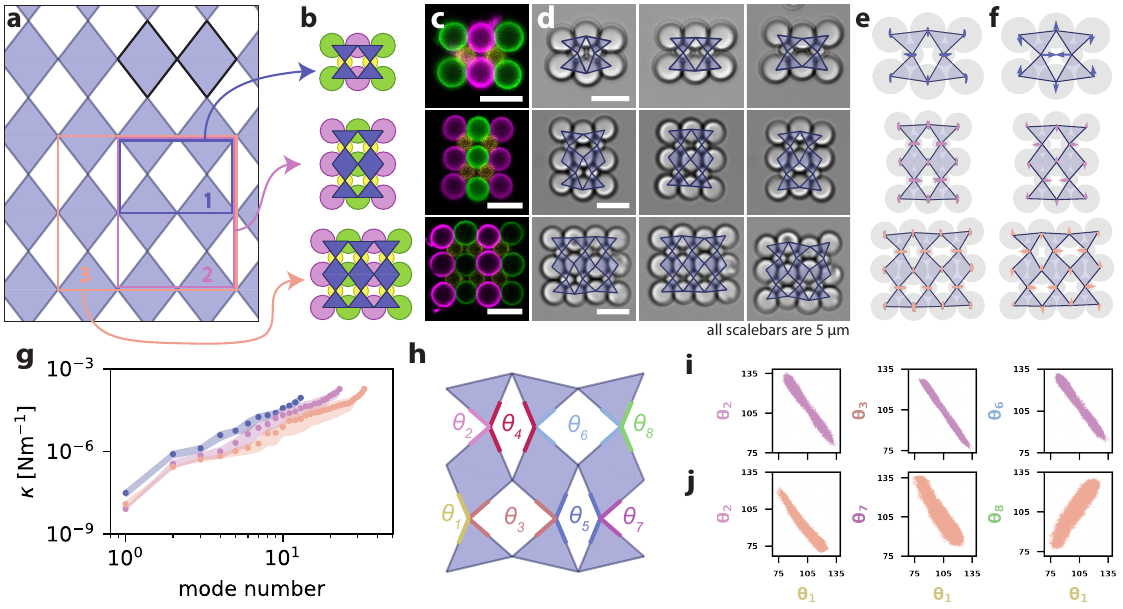}
    \caption{\textbf{A colloidal mechanism - the rotating diamond structure.} (a) The rotating diamond lattice where the hinging unit is indicated in black. Different subsets of the lattice were translated to the colloidal hinging system as schematically shown in (b). (c) Confocal images show the experimental realization where colors indicated different linker combinations. (d) Brightfield microscopy snapshots show the structural flexibility through thermal fluctuations, with the undeformed  state in the central column and examples of changes along opposite directions of the mode in the left and right columns. (e) Extracted positions over time corrected for global rotation and translation, which are used for mode analysis. (f) All three structures possess a single soft mode shown in (f), which is a counter-rotation of the rigid units. (g) This mode is an order of magnitude softer than the second-softest mode as shown in the mode spectrum. Correlations in the (h) different opening angles show excellent correlation and anti-correlation as exemplified for structures (i) 2 and (j) 3 evidencing the counter-rotating motion expected for auxetic behavior. Shaded regions in (g) represent the standard deviation.}
    \label{fig:fig2}
\end{figure*}

To achieve the crucial combination of freely hinging and non-spherical elements, we assemble silica spheres with surface-mobile DNA linkers into rigid, non-spherical clusters that pivot flexibly around a shared particle (Figs.~\ref{fig:fig1}b,c). The simplest case of a pair of triangular clusters requires two particles (with diameter $d=2.12\pm0.06~\mu$m) equipped with type A and B linkers, two with A' and B linkers, and one with only B' linkers, where accents denote complementary sequences (Fig.~\ref{fig:ext_dat_fig1}). Although all links in the resulting structure allow for sliding, our bond design restricts the motion to rotations and creates mechanically rigid triangular clusters of spheres. 
The remaining single degree of freedom is collective hinging of the triangles around the central B' particle, realizing a flexibly rotating colloidal pivot with a hinging point embedded within the central particle (Fig.~\ref{fig:fig1}b). 
Following this design principle, we furthermore create a 4-valent rigid diamond pivot. Diamonds of equal sized spheres form a square instead of a rhombus and thus cannot pivot due to steric hindrance (Fig.~\ref{fig:fig1}c). We therefore use larger ($d=3.34\pm0.14~\mu$m) particles with either AB or A'B type linkers together with smaller particles ($d=2.12\pm0.06~\mu$m) equipped with B' linkers only which maximize the angular range of motion (Methods and Fig.~\ref{fig:ext_dat_fig2}).
The linker design and orientation of the clusters ensure that neighboring units do not bond and their density and size lead to sedimentation and quasi-2D motion (Fig.~\ref{fig:ext_dat_fig1}).

Our colloidal pivots are so flexible that they freely hinge around their shared particle as evidenced by their thermal motion (Figs.~\ref{fig:fig1}c,d). Their relative angular motion is only limited by steric constraints, with relative angles ranging from $-60\degree$ to $60\degree$ for hinging triangles and from $-30\degree$ to $30\degree$ for hinging diamonds (Figs.~\ref{fig:fig1}e,f). The ultralow hinge stiffness therefore has an entropic origin and is of the order 10~nN/m (Fig.~\ref{fig:ext_dat_fig3} and Methods), which is four orders of magnitude softer than the most advanced previous  method~\cite{smart_magnetically_2024}. 
Our strategy thus combines the binding and sliding of multiple species of spherical colloidal joints to realize rigid, non-spherical elements which are flexibly hinged.
We first demonstrate how colloidal pivots allow to materialize metamaterials on the micron scale by creating the archetypical lattices of flexibly hinged quadrilaterals~\cite{Resch1963,Grima2000,mullin_pattern_2007} (Figs.~\ref{fig:fig2}a,b and Methods). 
We find that our hierarchically built structures fluctuate along a single soft counter-rotating deformation mode (Figs.~\ref{fig:fig2}c-f), agreeing with predictions for the underlying bond network~\cite{Maxwell1864,Lubensky2015}.  
The entropic stiffness of this mode is significantly smaller than that of all other modes, which require stretching or compression of bonds (Fig.~\ref{fig:fig2}g). Brownian particle simulations show excellent agreement with experiments and comparison with stochastic conformation sampling and a linear response model allows identification of contributions from entropy and the spring-like character of the bonds
(Methods and Fig.~\ref{fig:ext_dat_fig3}). 
The dominance of this single soft mode produces the characteristic counter-rotation of the clusters and the checkerboard pattern of holes~\cite{Resch1963,Grima2000,mullin_pattern_2007}.
Indeed, the internal angles of the structure show excellent (anti-)correlations 
(Figs.~\ref{fig:fig2}h-j, Fig.~\ref{fig:ext_dat_fig4} and Methods). 
These findings highlight that the microscopic rotating diamonds structure forms a dilational metamaterial which is activated by thermal fluctuations.

\begin{figure*}
    \centering
    \includegraphics[width=\textwidth]{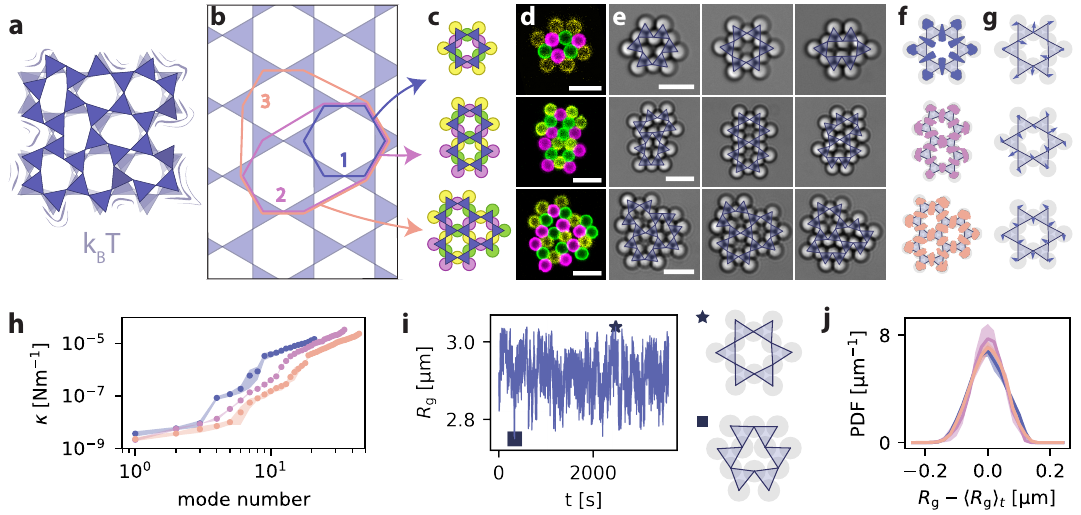}
    \caption{\textbf{Thermally actuated rotating triangle structures.} (a) A schematic representation of a floppy Kagome lattice. (b) The selection of subsets of the Kagome lattice which are translated to the colloidal hinging system as schematically shown in (c). (d) Confocal microscopy images show the experimental realization (different particle functionalizations are indicated by color and fluorescent dye) and brightfield microscopy snapshots in (e) show the structural flexibility by thermal fluctuations. (f) The global translation and rotation subtracted particle positions that are used for mode analysis. (g) The three softest modes for the smallest structure. (h) The mode spectrum. (i)  Time series of the radius of gyration for the smallest structure with the conformations with the maximum and minimum $R_g$ indicated. (j) Histogram of the radius of gyration $R_g(t)$ with respect to its time averaged value of $\langle R_g(t)\rangle _t$ from (i). Shaded regions in (h) and (j) represent the standard deviation.}
    \label{fig:fig3}
\end{figure*}

For creating devices and machines, larger motions and multiple reconfiguration possibilities, or soft modes, are desirable.
One such structure is the well-known rotating triangles, or Kagome, lattice, of which 
we create three different sublattices by arranging rigid triangles into 1, 2, and 3 connected rings (Figs.~\ref{fig:fig3}a-d). 
We find that these display large thermal fluctuations that lead to an opening and closing of the rings (Fig.~\ref{fig:fig3}e). The larger range of motion and multiple soft modes allow for more displacement freedom (Fig.~\ref{fig:fig3}f).

Quantitative analysis reveals that already the smallest Kagome structure possesses three soft modes, including the auxetic deformation of the idealized rotating triangle structure~\cite{Grima2000, Lubensky2015}, which all involve rotation of individual rigid triangles (Fig.~\ref{fig:fig3}g). 
In contrast to the rotating diamond structure, scaling these Kagome structures up has been predicted to lead to an increasing number of soft modes as $n_t-2n_r-1$, with $n_t$ the number of triangle units and $n_r$ the number of rings~\cite{Maxwell1864, sun_surface_2012}, in agreement with the mode spectra found in experiments (Fig.~\ref{fig:fig3}h), simulations, stochastic sampling of the conformations, and a linear response model (Fig.~\ref{fig:ext_dat_fig5}). 

To quantify their auxetic behavior, we measure their radius of gyration $R_\mathrm{g}$, which captures the mass distribution. $R_\mathrm{g}$ fluctuates strongly, with the largest value corresponding to the undeformed structure and the smallest to the state where all triangles have been fully rotated inwards. 
We find that our thermal Kagome structures most often reside at intermediate $R_g$ values, in between the fully extended and fully compressed state, where most microstates are available (Figs.~\ref{fig:fig3}i,j).  

\begin{figure*}
    \centering
    \includegraphics[width=\textwidth]{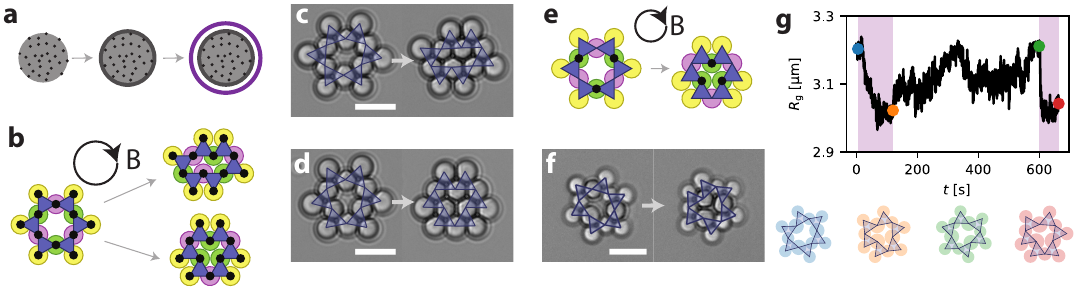}
    \caption{\textbf{Controllable deformation by magnetic actuation.} (a) To create magnetically responsive colloidal Kagome structures, colloids containing superparamagnetic iron oxide nanoparticles are coated with silica and functionalized with lipids and DNA linkers. (b) In a fast rotating magnetic field (Hz),  magnetic colloids attract, thereby compacting the Kagome ring in either of two ways, as experimentally observed in (c,d) brightfield microscopy. (e,f) When only three particles are magnetically responsive (indicated in green) they compress along the purely auxetic mode, as depicted in the e) schematic and (f) experimental realization. (g) When the field is turned on (purple), the radius of gyration $R_g$ decreases as the structure folds along the mode. When the field is off, thermal fluctuations unfold the structure again. Conformations at points indicated are shown below. 
    }
    \label{fig:fig4}
\end{figure*}

To create functionality in such a floppy structure, we integrate magnetic colloidal joints to control the deformation. We coat colloidal silica particles that contain superparamagnetic nanoparticles 
with a St\"ober silica layer to enable functionalization with lipids and DNA-linkers and assemble them in a Kagome ring (Fig.~\ref{fig:fig4}a and Methods). 
To realize effective attraction between the particles we use a fast-rotating magnetic field $\vec{B}$ in the observation plane. Upon application of the magnetic field, the Kagome ring folds into one of two conformations (Figs.~\ref{fig:fig4}b-d, Fig.~\ref{fig:ext_dat_fig6}, and Methods).

To favor deformations exclusively along the auxetic mode, we combine three magnetic and nine nonmagnetic particles into a Kagome structure (Figs.~\ref{fig:fig4}e,f). In a rotating magnetic field, only the magnetic particles attract and are pulled inwards, selectively actuating the auxetic deformation mode (Fig.~\ref{fig:fig4}f). The folding is accompanied by a decrease in $R_g(t)$ to its minimum and can be reversed by thermal fluctuations when the field is turned off (Figs.~\ref{fig:fig4}f,g). This Kagome-based colloidal mechanical metamaterial thus has an externally actuatable auxetic mode, making it a simple version of a Brownian machine that can fold by magnetic attractions and unfold by thermal fluctuations.

Our ability to create colloidal mechanisms and mechanical metamaterials from colloidal pivots opens the door to translate the large variety of macroscopic mechanical metamaterial designs to the microscale. 
Their controlled response to thermal fluctuations and magnetic fields provides great potential for use in mechanical devices that are able to sense, actuate, compute, and autonomously take decisions~\cite{yasuda_mechanical_2021, chen_reprogrammable_2021, ball_animate_2021,jiao_mechanical_2023, byun_integrated_2024, dudek_shape_2025}, and makes them exciting model systems to provide quantitative insights into how thermal fluctuations impact biological machines and materials~\cite{kolomeisky2007molecular, dennison_fluctuation-stabilized_2013, broedersz_modeling_2014, mannattil_thermal_2022}.

\renewcommand{\bibnumfmt}[1]{[M#1]}
\renewcommand{\citenumfont}[1]{M#1}

\newpage
\section*{Methods}
\textbf{Colloidal particles.}
Monodisperse silica particles with diameters $2.12\pm0.06$~$\mathrm{\mu m}$ and $3.34\pm 0.14$~$\mathrm{\mu m}$, as well as paramagnetic particles of a composite material of iron oxide nanoparticles dispersed in a silica matrix of size $2.54\pm0.09~\mathrm{\mu m}$ with an iron oxide content of at least 25\% (susceptibility $\chi=0.88$) were purchased from Microparticles GmbH and stored in water at a concentration of 5 wt\% at $4\degree~\mathrm{C}$.
The paramagnetic particles were coated with a thin layer of silica following a St\"ober protocol~\cite{Stober1968} before further use. For this, $250~\mathrm{\mu L}$ of the 5~wt\% stock particles was added to 5~mL of ethanol (96\%). Then, $750~\mathrm{\mu L}$ of ammonium hydroxide solution (NH$_3$ in water, 28.0-30.0\%, Honeywell) as well as $25~\mathrm{\mu L}$ of tetraethyl orthosilicate (TEOS, $\geq$~99\%, Sigma Aldrich) were added. The mixture was put in an ultrasonication bath for 5~hours in total. After 3~hours, another $25~\mathrm{\mu L}$ of TEOS was added and sonication was continued for 2h. Then, the particles were washed 2~times in ethanol and 2~times with water (Milli-Q IQ 7000, resistivity 18.2~M$\ohm$~cm at 25~$\degree$C, total organic carbon ~$<$~5~ppb and diluted to a concentration of approximately $5~\mathrm{gL^{-1}}$.

\textbf{DNA hybridization.} Single-stranded (ss) DNA strands were purchased from Eurogentec as custom sequences specified in Table~\ref{table:dna}. Complementary sticky end sequences are indicated by use of the same letters for naming the strands, with and without primes.
DNA was hybridized by diluting equimolar amounts of ssDNA in buffer (200~mM sodium chloride (NaCl, extra pure, Acros Organics), 10~mM 2-[4-(2-hydroxyethyl)-1-piperazinyl]ethanesulphonic acid (HEPES, $\geq 99.5\%$, Carl Roth), pH~7.40), heating it to 94 $\degree$C for 30 minutes in an oven (Memmert UFE 500), and allowing it to cool down slowly overnight by turning off the heating. Inert double stranded (ds) DNA used for steric stabilization was obtained by mixing strands ssI and ssI' to get dsI.
Linker dsDNA was obtained by mixing strands ssA ssA', ssB, or ssB' with the backbone strand ssBB to get dsDNA linkers A, A', B, or B'
with single-stranded sticky ends indicated in italic in Table \ref{table:dna}. 

\textbf{Particle functionalization and assembly.} The procedure to make colloidal joints was adapted from Ref.~\citenum{melio_soft_2024}: first, the colloids were coated with a lipid bilayer containing PEGylated lipids for steric stabilization and dyed lipids for identification purposes, into which, in a second step, DNA linkers and inert DNA strands were inserted. 
For assembling colloidal pivots, particles were functionalized with one or two types of DNA linkers and fluorescently labeled using one of three dyed lipids as illustrated in Figs.~\ref{fig:fig1}b,c to enable identification during and after assembly. As dyed lipids either 23-(dipyrrometheneboron difluoride)-24-norcholesterol (TopFluor® Cholesterol, 1 $\mathrm{gL^{-1}}$, Avanti Polar Lipids), 1,2-dioleoyl-sn-glycero-3-phosphoethanolamine-N-(lissamine rhodamine B sulfonyl) (ammonium salt) (DOPE-Rhodamine, 1 $\mathrm{gL^{-1}}$, Avanti Polar Lipids), or Atto390-labeled DOPE (1 $\mathrm{gL^{-1}}$, ATTO-TEC GmbH) were used. Lipids were dissolved in chloroform (99.0 - 99.4\%, Sigma Aldrich) at specified concentrations and stored at -18~$\degree$C.

Particles were assembled into colloidal pivots by bringing them in contact using a custom-built optical tweezer setup (1064 nm laser, Laser QUANTUM). Particle interactions were designed such that triangles and diamonds assembled following the designs laid out in Figs.~\ref{fig:fig1}b,c were mechanically stable~\cite{Maxwell1864} yet were able to hinge around shared particles (see Fig.~\ref{fig:ext_dat_fig1}).

\textbf{Data acquisition.} The samples were imaged with an inverted confocal microscope (Nikon Ti-E) equipped with a Nikon A1R confocal scanhead with Galvano and resonant scanning mirrors. A 60$\times$ water immersion objective (NA = 1.2) was used to image the sample. Lasers of 405~nm, 488~nm, and 561~nm wavelength were used to excite the Atto390, TopFluor-Cholesterol and Rhodamine dyes, respectively. The excitation light was passed through a quarter wave plate to prevent polarization of the dyes and the emitted light was separated by 425-475~nm, 500–550~nm, and 565–625~nm filters to identify a different DNA functionalization in each channel.

The particle positions were obtained from brightfield microscopy videos taken at a frame rate of 5-20 fps using a least-square fit of a Mie scattering based model implemented in python package HoloPy~\cite{Barkley2020,Verweij2023,melio_soft_2024}.
The size of a pixel corresponds to 110~nm and the sub-pixel tracking resolution was found to be on the order of 10-20~nm~\cite{melio_soft_2024}.
For rotating diamond and triangle structures 30 mins of video data from at least two structures were analyzed.

\textbf{Subtraction of trivial modes.} As we are interested in the internal reconfigurations only, we subtract translational and rotational displacements of the structure as a whole from the motion of the particles. The translational diffusion of the structures is subtracted by placing the origin of the tracking frame in the center of mass of the networks, where we assume particles having the same mass. For particle $i$, the translation corrected position at time $t$ is then given by
\begin{equation}
    \vec{r}_{i,\mathrm{trans}}(t) = \vec{r}_{i}(t)
    - \frac{1}{N}\sum_{j=1}^N
    \vec{r}_{j}(t)
    ,
    \label{eq:translation_correction}
\end{equation}
\noindent where $\vec{r}_{i}(t)$ is the tracked position and $N$ is number of particles.

To subtract rotation of the structures, we calculated the average angle of all bonds with respect to the $x$-axis of the lab frame. For a bond between particles $i$ and $j$, the angle with the lab frame is given by $\theta_{ij}=\arctan{\Big(\frac{y_{j}-y_{i}}{x_{j}-x_{i}}\Big)}$. The average angle is then calculated as $\theta_\text{avg}=\frac{1}{2n_\text{b}}\sum_i^N\sum_j^N \theta_{ij}$, summed over all pairs of particles $i$ and $j$ that share a bond, and where $n_\text{b}$ is the total number of bonds. The rotation corrected position is then calculated as a rotation by this mean angle,
\begin{equation}
    \vec{r}_{i,\mathrm{rt}}(t) = \vec{r}_{i,\mathrm{trans}}(t)
    \begin{bmatrix}
    \cos{\theta_\text{avg}(t)}& \sin{\theta_\text{avg}(t)}\\
    -\sin{\theta_\text{avg}(t)}&\cos{\theta_\text{avg}(t)}
    \end{bmatrix},
    \label{eq:rotation_correction}
\end{equation}
which defines the directions of the $x$- and $y$-axes as $\vec{x} = \vec{r}_{i,\mathrm{rt,x}}$ and $\vec{y} = \vec{r}_{i,\mathrm{rt,y}}$.

\textbf{Analysis of pivoting motion.} We extract information about the thermal fluctuation of a single pivot made from two triangular or diamond-shaped structures by tracking the positions of all particles (Fig.~\ref{fig:fig1}f). We measure the opening angle between the two neighboring particles connected to the hinging pivot as $\theta$ (Figs.~\ref{fig:fig1}d,e), where the undeformed state is assigned a value $\theta_0$, corresponding to $120\degree$ and $75\degree$ for the triangular pivots and diamond pivots, respectively. Fluctuations in this opening angle are constrained to the geometrically possible range, as shown in Figs.~\ref{fig:fig1}d-f. 

We can get an estimate for the emergent entropic stiffness $k_{NNN}$ of our colloidal pivots by measuring the probability distributions of opening angles $P(\theta)$. If we approximate this distribution with a normal distribution of the same variance, we obtain an angle potential $V=\frac{1}{2}k_\theta(\theta-\theta_0)^2$ with $k_\theta$ an angle spring that is related to the angle variance as $k_\theta=\frac{k_\text{B}T}{\left\langle(\theta-\theta_0)^2\right\rangle}$. To get the spring in the right units, we take the second derivative to the next-nearest-neighbor distance given by $x=d(1+\cos{\theta})$ with $d$ the unit side length distance, which is $d=2r$ for triangular units of particles with radius $r$, and $d=r_s+r_b$ for diamond-shaped units with particle radii $r_s$ and $r_b$. This results in
\begin{equation}
    \frac{\partial^2V}{\partial x^2} = k_\theta\left(\frac{\partial\theta}{\partial x}\right)^2+k_\theta(\theta-\theta_0)\frac{\partial^2\theta}{\partial x^2} \quad,
\end{equation}
which we evaluate around $\theta_0$ resulting in
\begin{equation}
    \frac{\partial^2V}{\partial x^2}\Bigg|_{\theta=\theta_0} = k_\theta\left(\frac{\partial\theta}{\partial x}\right)^2 = \frac{k_\theta}{d^2\left(1-(\frac{x|_{\theta=\theta_0}}{d}-1)^2\right)} \quad.
\end{equation}
For the triangular pivots, $x|_{\theta=120\degree}=3r$ leads to $k_{NNN}=\frac{k_\theta}{3r^2}$. From the angle distribution we measure a variance of 0.203 rad$^2$, which results in $k_{NNN}=6.0~\mathrm{nNm^{-1}}$. For diamond-shaped pivots, $x|_{\theta\approx75\degree}=2\arccos{\frac{r_b}{r_s+r_b}}$. From the angle distribution we measure 0.041~rad$^{-1}$ which results in $k_{NNN}=15.1~\mathrm{nNm^{-1}}$

\textbf{Brownian particle simulations.}
Brownian particle simulations are performed with bonds between particles modeled as springs with spring constant $k_\text{bond}=32~\mathrm{\mu Nm^{-1}}$ and an additional one-sided repulsive spring of stiffness $k_\text{bond}$ acts whenever non-bound particles overlap~\cite{melio_soft_2024}.
Particles are initialized on undeformed rotating diamonds and Kagome lattices. The overdamped Langevin equation is solved using a first-order Euler method with a simulation time step $dt=20~\mathrm{\mu s}$ and particle mobility $\mu=\frac{1}{6\pi\eta r}$ with $\eta$ the viscosity and $r$ the particle radius. The thus obtained position vectors $\vec{r}(t)$ are corrected for translations and rotations by Eqs.~\ref{eq:translation_correction}~and~\ref{eq:rotation_correction}.

\textbf{Stochastic conformation sampling}
To identify contributions to the modes stemming from floppy reconfigurations only, we employ a Monte-Carlo approach to sample all conformations assuming  fixed bond lengths. This is justified as particle bond fluctuations are small (10-50~nm) in comparison to the displacements along floppy modes (1-2~$\mu$m). 
For this, we parameterize the particle positions using opening angles and draw random values from a uniform distribution for each degree of freedom. A single opening angle is sufficient for describing the rotating diamonds structure, while $n_t-2n_r-1$ opening angles are needed for Kagome structures with $n_t$ the number of triangle units and $n_r$ the number of rings. Only combinations of opening angles that do not result in overlapping particles are saved, and the resulting position vectors $\vec{r}(\vec{\theta})$ are used for mode analysis.

\textbf{Analytical linear spring model.} To investigate the effect of the spring-like interactions between bonded particles on the mode analysis, we construct $2N\times2N$ stiffness matrices $\mathbf{K}$ for structures consisting of $N$ particles containing all spring interactions between the $2N$ particle coordinates.
The interaction between two particles $i$ and $j$ results in an off-diagonal stiffness matrix block of
\begin{equation}
    \mathbf{K}_{ij} = -k_\text{bond} 
    \begin{bmatrix}
        (\hat{b}\cdot\hat{x})^2 & (\hat{b}\cdot\hat{y})(\hat{b}\cdot\hat{x})\\
        (\hat{b}\cdot\hat{x})(\hat{b}\cdot\hat{y}) & (\hat{b}\cdot\hat{y})^2
    \end{bmatrix} ,
\end{equation}
where $\hat{b}$ is the unit vector along the interaction and $\hat{x}$ and $\hat{y}$ are unit vectors in $x$- and $y$-direction. Then, by Newton's third law, the diagonal stiffness matrix block for particle $i$ is
\begin{equation}
\mathbf{K}_{ii} = -\sum_i^N\mathbf{K}_{ij}, i\neq j .
\end{equation}
The eigensystem of this matrix gives the modes $\vec{\xi}_\nu$ and stiffnesses $\kappa_\nu$ directly.\\

\textbf{Mode analysis.} For analysis of the modes, we assume linear response, such that the displacement of each particle from its equilibrium position results in a restoring spring force back to equilibrium. We obtain the modes and accompanying stiffnesses by calculating the covariance matrix of the particle displacements from their equilibrium positions $\mathbf{C_p}$, which for sufficiently long measurements and in linear response are related to the stiffness matrix by $\mathbf{K}=\frac{k_\mathrm{B}T}{\mathbf{C_p}}$~\cite{Henkes}. The eigenvectors of $\mathbf{K}$ are the modes $\vec{\xi}_\nu$ and the eigenvalues are the related stiffnesses $\kappa_\nu$.

In experiments and simulations, we obtain these displacements by measuring the particle positions $\vec{r}(t)$ over a long measurement time. For sufficiently long times, the equilibrium position is equal to the average position $\langle\vec{r}(t)\rangle_t$ where $\langle\dots\rangle_t$ denotes a time-average. The displacement vector can then be defined as $\delta\vec{r}(t)=\vec{r}(t)-\langle\vec{r}(t)\rangle_t$. 
For the stochastic conformation sampling approach, an opening angle dependent position vector is used instead, such that $\delta\vec{r}(\vec{\theta})=\vec{r}(\vec{\theta})-\langle \vec{r}(\vec{\theta})\rangle _{\vec{\theta}}$ with $\langle\dots\rangle_{\vec{\theta}}$ denoting an average over $\vec{\theta}$.
The covariance matrix $\mathbf{C_p}$ of either displacement vector is calculated as $\mathbf{C_p} = \langle\delta\vec{r}\:\delta \vec{r}^T\rangle_t$. The stiffness matrix is obtained by a pseudo-inversion of $\mathbf{C_p}$ as $\mathbf{K} = k_\mathrm{B}T\mathbf{C_p}^{-1}$, which is necessary since $\mathbf{C_p}$ is singular as a result from subtracting global translations and rotations. The eigensystem of $\mathbf{K}$ contains the modes $\vec{\xi}_\nu$ and stiffnesses $\kappa_\nu$.
The experimentally obtained mode spectra for the rotating diamonds and rotating triangles structures are shown in Fig. \ref{fig:ext_dat_fig2}g and \ref{fig:fig3}h, respectively. A comparison between the mode spectra obtained from experiments, Brownian dynamics simulations, the stochastic approach and the spring model are presented in Figs.~\ref{fig:ext_dat_fig4}, \ref{fig:ext_dat_fig5}.

\textbf{Radius of gyration.}
We calculate the radius of gyration $R_g$ shown in Figs.~\ref{fig:fig3}i,j and \ref{fig:fig4}g  from the positions of all $N$ particles as
\begin{equation}
    R_g = \sqrt{\frac{1}{N}\sum_i^N{|\vec{r}_i|^2}} \quad,
\end{equation}
where $\vec{r}_i$ is the position with respect to the center of mass of the structure.

\textbf{Magnetic actuation experiments.} The magnetic coil setup used to generate magnetic fields with vector parallel to the observation plane consists of two coil pairs to apply magnetic fields in the $x$-direction and $y$-direction. Each coil consists of copper wire (diameter 0.3~mm) with an approximate winding number of 650. Both the $x$-coil pair and $y$-coil pair are connected to a custom Howland current amplifier which is connected to a function generator (Agilent 33500B series).
A 2~Volt peak to peak input signal as used in the here presented experiments corresponds to a magnetic field of about $10^3~\mathrm{Am^{-1}}$.

Upon applying an external magnetic field $\vec{B}$, the superparamagnetic particles form a magnetic dipole moment $\vec{m}$, which we approximate as a point dipole, see Fig.~\ref{fig:ext_dat_fig6}a.
If the rotation frequency of the external magnetic field is sufficiently high such that the field rotates faster than the particles reposition, the potential energy $V$ and force $F$ between any two particles can be replaced with their time-averages as
\begin{equation}
    \begin{split}
        \langle V_\text{dipoles}\rangle &= -\frac{\mu_0m^2}{8\pi r^3}\\
        \langle\vec{F}\rangle &= -\frac{3\mu_0m^2}{8\pi r^4}\hat{e}_r
    \end{split} ,
    \label{eq:time_average_magnetic_field}
\end{equation}
with $\mu_0$ the vacuum permeability, $r$ the center-to-center distance between the particles, and $\hat{e}_r$ the unit vector along their bond. The criterion that the repositioning dynamics of the particles are much slower than the rotation frequency of the external field is met for frequencies greater than  $1-100~\mathrm{mHz}$, at which the magnetic structures could still follow the rotation of the field. Magnetic field rotation frequencies of $10~\mathrm{Hz}$ were thus used in experiments.

Using the expression for the time-averaged potential energy, we can calculate the contribution from the potential energy induced by the magnetic interaction to the free energy for all conformations. For Kagome structures made of 12 superparamagnetic particles, eight energy minima are found (Figs.~\ref{fig:ext_dat_fig6}c,d), which correspond to two distinguishable conformations, as were also observed in experiments (Fig.~\ref{fig:fig4}). By changing the design using only three superparamagnetic particles, only one potential energy minimum is found and corresponds to the conformation observed in experiments (Figs.~\ref{fig:ext_dat_fig6}f,g).

\textbf{Design of the rotating diamond structure.}
\label{sec:RotDiamond}
To build a rotating diamond structure that can be deformed by thermal fluctuations, unencumbered hinging is required. The only particle shape that provides perfect hinging are spheres, because anisotropic particle shapes lead to limited flexibility due to the surface-mobility of the DNA linkers and multivalent character of the bond~\cite{chakraborty_colloidal_2017,  linneDirectVisualizationSuperselective2021,shelke_flexible_2023}.
We thus realized thermally flexible rotating diamonds structures by placing a sphere at each hinging point or corner of the diamond unit, see Fig.~\ref{fig:ext_dat_fig2}a. In addition, we employed two types of spheres with radii $r_b$ and $r_s$ as depicted in Fig.~\ref{fig:fig1}c, where the indices $b$ and $s$ refer to big and small particles, respectively, to allow the counter-rotating motion of neighboring diamond units.
The range of motion depends on the size ratio of the smaller and larger spheres, $r_s/r_b$. 
For $r_s/r_b \Rightarrow 1$, hexagonal close packing of the particles prevents motion, see Fig.~\ref{fig:ext_dat_fig2}b. 
For $r_s/r_b \Rightarrow \sqrt{2}-1$, which corresponds to a rotating squares mechanism, the bigger spheres touch, thereby rigidifying the mechanism as well, see Fig.~\ref{fig:ext_dat_fig2}c. 
At intermediate values for $r_s/r_b$, motion is possible for which we quantify the range using the conformations obtained from Monte Carlo sampling and identifying the angles that are geometrically allowed for a given size ratio $r_s/r_b$. The resulting range of allowed angles is shown in Fig.~\ref{fig:ext_dat_fig2}e. The optimum is found at $r_s/r_b=0.638$ with steep decays towards both higher and lower size ratio's. 

We can derive the maximum value for the range of motion analytically, by considering that the two types of inner angles of the holes, $\theta_1$ and $\theta_2$ are related by $\theta_1+\theta_2=2\alpha$, where $\alpha=2\sin^{-1}\big(\frac{r_b}{r_b+r_s}\big)$, as indicated in Fig.~\ref{fig:ext_dat_fig2}a~\cite{Resch1963,Grima2000}.
For both angles, the minimum value $\theta_{\text{min}}$ is obtained upon contact of the small spheres, i.e. for $\theta_{\text{min}}=2\sin^{-1}\big(\frac{r_s}{r_b+r_s}\big)$, see Fig.~\ref{fig:ext_dat_fig2}d. Similarly, their maximum value $\theta_\text{max}$ is limited by contact of the big spheres touching, i.e. $\theta_\text{max}=2\cos^{-1}\big(\frac{r_b}{r_b+r_s}\big)$. 
These constraints together with the constraint on the sum of two neighboring inner angles yields for the maximally rotated case, i.e. for $\theta_1=\theta_\text{min}$ and $\theta_2=\theta_\text{max}$,
\begin{equation}
    \begin{split}
        \theta_1+\theta_2=\theta_\text{min}+\theta_\text{max}&=2\alpha\\
        \rightarrow 2\sin^{-1}\left(\frac{r_s}{r_b+r_s}\right)+2\cos^{-1}\left(\frac{r_b}{r_b+r_s}\right)&=4\sin^{-1}\left(\frac{r_b}{r_b+r_s}\right),
    \end{split}
\end{equation}
This equation can be solved numerically to yield a ratio of $r_s/r_b=0.638$.

To maximize the range of motion in experiments, we used particles with $r_s=1.06~\mathrm{\mu m}$ and $r_b=1.67~\mathrm{\mu m}$ yielding a ratio of $r_s/r_b=0.635$, which is very close to the optimum value for the maximum range of motion.

\textbf{Correlated and anticorrelated angles in rotating diamond structures.}\label{sec:angle_correlations}
The rigid units of rotating diamond structures are connected such that their single soft mode manifests itself by counter-rotating neighboring units~\cite{Maxwell1864,Resch1963,Calladine1978,Grima2000,Lubensky2015}. This implies that opening angles of neighboring units and units separated by an odd number of units are expected to be perfectly anticorrelated and units separated by an even number are perfectly correlated.
We investigate the correlations and anticorrelations of different opening angles with respect to one given angle $\theta_1$ in the rotating diamond structure in Fig.~\ref{fig:ext_dat_fig3}. In panel a, different angles are indicated and their correlations are plotted in panels (b-h). 
To assess the quality of the correlation, we fit them with a straight line as $\theta_i=a\theta_1+b$ with $a$ the slope, and report the value of $a$ in the inset. For perfect correlations, $a=1$, and for perfect anticorrelations, $a=-1$ is expected. 

For opening angles $\theta_1$ and $\theta_2$, separated by one unit in the $y$-direction, a clear anticorrelation is found with a slope of $a=-0.95$, which is close to $-1$ (Fig.~\ref{fig:ext_dat_fig3}b). For opening angles $\theta_4$ and $\theta_1$, separated by two units, one in $x$-direction and one in $y$-direction, the correlation is $a=1.07$, close to $+1$ (Fig.~\ref{fig:ext_dat_fig3}c). 
For one unit yet further away, opening angles $\theta_1$ and $\theta_6$ show an anti-correlation (Fig.~\ref{fig:ext_dat_fig3}d), while the furthest separation distance, $\theta_1$ and $\theta_8$, is correlated again (Fig.~\ref{fig:ext_dat_fig3}e). The same checkerboard pattern of correlations and anticorrelations is observed comparing opening angles in $x$-direction as shown in Figs.~\ref{fig:ext_dat_fig3}f-h.
These near-perfect correlations and anti-correlations underline that the colloidal rotating diamonds structure is a mechanism.

Correlations between angles separated by a larger distance furthermore have a higher variance. To identify the cause of this, we first note that the bond length between small and big particles fluctuates more than between two larger particles as illustrated in Figs.~\ref{fig:ext_dat_fig3}e and j.
The distribution of the bond length variation between two big particles (purple symbols, Fig.~\ref{fig:ext_dat_fig3}l) is quite narrow as exemplified by a standard deviation of $\sigma=23~\mathrm{nm}$ for fitting with a normal distribution, similar to what was previously observed~\cite{melio_soft_2024}. However, the distribution of the bond length variation between a big and a small particle (green squares, Fig.~\ref{fig:ext_dat_fig3}l) is significantly wider, with a standard deviation of  $\sigma=51~\mathrm{nm}$. We attribute the wider distribution to out-of-plane motion against gravity (in the $z$-direction), which is greater for smaller particles. Any $z$-motion of smaller spheres changes the observed bond length, as brightfield microscopy observations are effectively restricted to projections onto the 2D plane. The difference in observed bond length can vary by up to 60~nm between the cases where all particles touch the substrate and where they have their centers at the same distance from the substrate. The $z$-displacement necessary for this is $0.61~\mu\text{m}$, 
significantly larger than the gravitational height of the small particle, which is only $h_s=(k_\text{B}T)/(m_sg)=45~\mathrm{nm}$. The motion of connected particles may help in lifting the smaller spheres, and other factors such as tracking uncertainty, stretching of the lipid bilayer and polydispersity in the colloidal shape might also contribute to the bond length fluctuations.

\textbf{Particle positions in the undeformed state.}\label{sec:undeformed_structures}
The undeformed states of the rotating diamonds and rotating triangles structures are used in simulations as initial positions and in the linear spring model to construct the stiffness matrix. The initial particle position vector is constructed as $\vec{r}=(x_1,y_1,\dots,x_N,y_N)$.

\begin{figure}
    \centering
    \includegraphics[width=\linewidth]{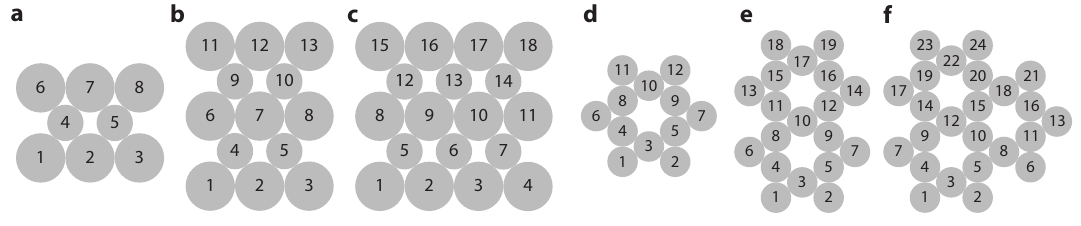}
    \caption{\textit{Particle numbering.} For rotating diamond structures with (a) $N=8$, (b) $N=13$, and (c) $N=18$, and Kagome structures with (d) $N=12$, (e) $N=19$, and (f) $N=24$.}
    \label{fig:particle_numbering}
\end{figure}

\textit{Rotating diamond structures:} Labeling the particles as depicted in Fig.~\ref{fig:particle_numbering}, the initial positions of a rotating diamond structure with $N=8$ and particle radii of $r_s$ and $r_b$ are given by
\begin{align}
 \vec{r}_{N=8} =\: &(0,0,2r_b,0,4r_b,0,r_b,\sqrt{r_s(r_s+r_b)}, \nonumber\\
 &3r_b,\sqrt{r_s(r_s+r_b)},0,2\sqrt{r_s(r_s+r_b)},\\
 &2r_b,2\sqrt{r_s(r_s+r_b)},4r_b,2\sqrt{r_s(r_s+r_b)}). \nonumber
\end{align}

For a rotating diamond structure with $N=13$, the position vector is given by

\begin{align}
 \vec{r}_{N=13} =\: &(0,0,2r_b,0,4r_b,0,r_b,\sqrt{r_s(r_s+r_b)},3r_b,\sqrt{r_s(r_s+r_b)}, \nonumber\\
 &0,2\sqrt{r_s(r_s+r_b)},2r_b,2\sqrt{r_s(r_s+r_b)},4r_b,2\sqrt{r_s(r_s+r_b)},\\
 &r_b,3\sqrt{r_s(r_s+r_b)},3r_b,3\sqrt{r_s(r_s+r_b)},0,4\sqrt{r_s(r_s+r_b)},\nonumber\\
 &2r_b,4\sqrt{r_s(r_s+r_b)},4r_b,4\sqrt{r_s(r_s+r_b)}). \nonumber
\end{align}

For a rotating diamond structure with $N=18$, the position vector is given by

\begin{align}
 \vec{r}_{N=18} =\: &(0,0,2r_b,0,4r_b,0,6r_b,0,r_b,\sqrt{r_s(r_s+r_b)},3r_b,\sqrt{r_s(r_s+r_b)}, \nonumber\\
 &5r_b,\sqrt{r_s(r_s+r_b)},0,2\sqrt{r_s(r_s+r_b)},2r_b,2\sqrt{r_s(r_s+r_b)},\nonumber\\
 &4r_b,2\sqrt{r_s(r_s+r_b)},6r_b,2\sqrt{r_s(r_s+r_b)},r_b,3\sqrt{r_s(r_s+r_b)},\\
 &3r_b,3\sqrt{r_s(r_s+r_b)},5r_b,3\sqrt{r_s(r_s+r_b)},0,4\sqrt{r_s(r_s+r_b)},\nonumber\\
 &2r_b,4\sqrt{r_s(r_s+r_b)},4r_b,4\sqrt{r_s(r_s+r_b)},6r_b,4\sqrt{r_s(r_s+r_b)}). \nonumber
\end{align}

\textit{Rotating triangle structures:}
Similarly, for a Kagome structure with $N=12$ and particle radii $r$, the initial position vector for particles labeled as in Fig.~\ref{fig:particle_numbering} is given by
\begin{align}
\vec{r}_{N=12}=\:&(-\sqrt{3}r,-3r,\sqrt{3}r,-3r,0,-2r,-\sqrt{3}r, \nonumber\\
&-r,\sqrt{3}r,-r,-2\sqrt{3}r,0,2\sqrt{3}r,0,-\sqrt{3}r,r,\\
&\sqrt{3}r,r,0,2r,-\sqrt{3}r,3r,\sqrt{3}r,3r). \nonumber
\end{align}

For a Kagome structure with $N=19$, the position vector is given by
\begin{align}
\vec{r}_{N=19}=\:&(-\sqrt{3}r,-5r,\sqrt{3}r,-5r,0,-4r,-\sqrt{3}r,-3r,\sqrt{3}r,-3r,-2\sqrt{3}r, \nonumber\\
&-2r,2\sqrt{3}r,-2r,-\sqrt{3}r,-r,\sqrt{3}r,-r,0,0,-\sqrt{3}r,r,\sqrt{3}r,r,\\
&-2\sqrt{3}r,2r,2\sqrt{3}r,2r,-\sqrt{3}r,3r,\sqrt{3}r,3r,0,4r,-\sqrt{3}r,5r,\sqrt{3}r,5r). \nonumber
\end{align}

And for a Kagome structure with $N=24$, the position vector is given by
\begin{align}
\vec{r}_{N=24}=\:&(-\sqrt{3}r,-5r,\sqrt{3}r,-5r,0,-4r,-\sqrt{3}r,-3r,\sqrt{3}r,-3r,3\sqrt{3}r, \nonumber\\
&-3r,-2\sqrt{3}r,-2r,2\sqrt{3}r,-2r,-\sqrt{3}r,-r,\sqrt{3}r,-r,3\sqrt{3}r,-r,\\
&0,0,4\sqrt{3}r,0,-\sqrt{3}r,r,\sqrt{3}r,r,3\sqrt{3}r,r,-2\sqrt{3}r,2r,2\sqrt{3}r,2r, \nonumber\\
&-\sqrt{3}r,3r,\sqrt{3}r,3r,3\sqrt{3}r,3r,0,4r,-\sqrt{3}r,5r,\sqrt{3}r,5r). \nonumber
\end{align}

\textbf{Parametrization using opening angles.}\label{monte-carlo_approach}
\textit{Rotating diamond structures.} For the rotating diamond structures, there is only one global reconfiguration mode. With the assumption of fixed bond lengths, the position of every particle can be described as function of a single opening angle. For the structure with $N=8$ where the particles are labeled as in Fig.~\ref{fig:particle_numbering}a, the positions as a function of opening angle $\theta$ are given by
\begin{equation}
    \begin{split}
        \begin{aligned}
            \vec{r}_1 &= \left(\begin{matrix}0\\0\end{matrix}\right)\\
            \vec{r}_2 &=2r_b \left(\begin{matrix}0\\1\end{matrix}\right)\\
            \vec{r}_3 &= \vec{r}_2+2r_b\left(\begin{matrix}\cos{(\frac{3}{2}\pi-\alpha-\theta)}\\\sin{(\frac{3}{2}\pi-\alpha-\theta)}\end{matrix}\right)\\
            \vec{r}_4 &= \left(\begin{matrix}\sqrt{r_s^2+2r_sr_b}\\r_b\end{matrix}\right)\\
        \end{aligned}
        \quad\quad\quad
        \begin{aligned}
            \vec{r}_5 &= \vec{r}_2+(r_s+r_b)\left(\begin{matrix}\cos{(\pi-\theta-\frac{\alpha}{2})}\\\sin{(\pi-\theta-\frac{\alpha}{2})}\end{matrix}\right)\\
            \vec{r}_6 &= \vec{r}_4+(r_s+r_b)\left(\begin{matrix}\cos{(\pi-\frac{3}{2}\alpha-\theta)}\\\sin{(\pi-\frac{3}{2}\alpha-\theta)}\end{matrix}\right)\\
            \vec{r}_7 &= \vec{r}_4+(r_s+r_b)\left(\begin{matrix}\cos{(-\theta+\pi-\frac{\alpha}{2})}\\\sin{(-\theta+\pi-\frac{\alpha}{2})}\end{matrix}\right)\\
            \vec{r}_8 &= \vec{r}_5+2r_b\left(\begin{matrix}\cos{(\frac{\pi}{2})}\\\sin{(\frac{\pi}{2})}\end{matrix}\right)
            \end{aligned}
    \end{split}
    \label{eq:geometric_rotating_diamond}
\end{equation}
where $\alpha=2\sin^{-1}{\left(\frac{r_b}{r_s+r_b}\right)}$. To obtain sample positions $\vec{r}(\theta)$, we use Eq. \ref{eq:geometric_rotating_diamond} with $\theta$ drawn from a uniform distribution between $\theta_\text{min}$ and $\theta_\text{max}$.

\begin{figure}[t]
    \centering
    \includegraphics[width=0.5\textwidth]{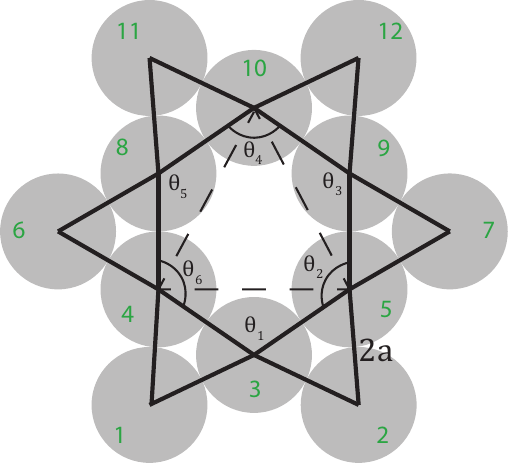}
    \caption{\textbf{Schematic of a Kagome structure} with particles in grey and bonds in black. The bond length is $2a$ and opening angles $\theta_i$ are indicated. The particles are labeled in green.}
    \label{fig:si_fig1}
\end{figure}

\textit{Rotating triangle structures.} 
For the Kagome structures, there are multiple global reconfiguration modes. The smallest structure with $N=12$ can be fully described with three opening angles $\theta_1$, $\theta_2$, and $\theta_3$ as indicated in Fig.~\ref{fig:si_fig1}. For sampling all conformations using a stochastic Monte-Carlo based approach, we draw random numbers for $\theta_1$, $\theta_3$, and $\theta_5$ from a uniform distribution between $\frac{\pi}{3}$ and $\pi$, set by the geometric limits. Then, angle combinations that are sterically allowed, i.e. do not posses overlapping particles, are selected. This is done by verifying whether the opening angles $\theta_2$, $\theta_4$, and $\theta_6$ are within the range of $\frac{\pi}{3}$ and $\pi$. For example, $\theta_4$ in Fig.~\ref{fig:si_fig1} is given by $\theta_4=\frac{\pi-\theta_3}{2}+\frac{\pi-\theta_5}{2}+\cos^{-1}{\left(\frac{\sin^2{(\frac{\theta_3}{2})}+\sin^2{(\frac{\theta_5}{2})}-\sin^2{(\frac{\theta_1}{2})}}{2\sin{(\frac{\theta_3}{2})}\sin{(\frac{\theta_5}{2})}}\right)}$. From the set of randomly sampled and sterically allowed angles, the particle positions can be calculated. For a Kagome structure with $N=12$ and bond lengths equal to $2r$, the following relations were used
\begin{equation}
    \begin{split}
        \begin{aligned}
            \vec{r}_1 &= 2r\left(\begin{matrix}0\\0\end{matrix}\right)\\
            \vec{r}_2 &= 2r\left(\begin{matrix}\frac{1}{2}+\sin{(\frac{\pi}{2}-\theta_1)}\\ \frac{1}{2}\sqrt{3}+\cos{(\frac{\pi}{2}-\theta_1)}\end{matrix}\right)\\
            \vec{r}_3 &= 2r\left(\begin{matrix}\frac{1}{2}\\ \frac{1}{2}\sqrt{3}\end{matrix}\right)\\
            \vec{r}_4 &= 2r\left(\begin{matrix}1\\0\end{matrix}\right)\\
            \vec{r}_5 &= 2r\left(\begin{matrix}\frac{1}{2}+\sin{(\frac{5}{6}\pi-\theta_1)}\\ \frac{1}{2}\sqrt{3}+\cos{(\frac{5}{6}\pi-\theta_1)}\end{matrix}\right)\\
            \vec{r}_6 &= 2r\left(\begin{matrix}1+\sin{(\frac{\pi}{6}+\theta_6)}\\ \cos{(\frac{\pi}{6}+\theta_6)}\end{matrix}\right)\\
        \end{aligned}
        \quad\quad\quad
        \begin{aligned}
            \vec{r}_7 &= \vec{r}_5+2r\left(\begin{matrix}\sin{(\frac{3}{2}\pi-\theta_2)}\\ \cos{(\frac{3}{2}\pi-\theta_2)}\end{matrix}\right)\\
            \vec{r}_8 &= \vec{r}_4+2r\left(\begin{matrix}\sin{(-\frac{\pi}{6}+\theta_6)}\\\cos{(-\frac{\pi}{6}+\theta_6)}\end{matrix}\right)\\
            \vec{r}_9 &= \vec{r}_5+2r\left(\begin{matrix}\sin{(\frac{11}{6}\pi-\theta_1-\theta_2)}\\ \cos{(\frac{11}{6}\pi-\theta_1-\theta_2)}\end{matrix}\right)\\
            \vec{r}_{10} &= \vec{r}_5+2r\left(\begin{matrix}\sin{(-\frac{7}{6}\pi+\theta_5+\theta_6)}\\ \cos{(-\frac{7}{6}\pi+\theta_5+\theta_6)}\end{matrix}\right)\\
            \vec{r}_{11} &= \vec{r}_4+2r\left(\begin{matrix}\sin{(-\frac{5}{6}\pi+\theta_5+\theta_6)}\\ \sin{(-\frac{5}{6}\pi+\theta_5+\theta_6)}\end{matrix}\right)\\
            \vec{r}_{12} &= \vec{r}_9+2r\left(\begin{matrix}\sin{(\frac{\pi}{2}-\theta_1-\theta_2-\theta_3)}\\ \cos{(\frac{\pi}{2}-\theta_1-\theta_2-\theta_3)}\end{matrix}\right)\\
        \end{aligned}
    \end{split}
\end{equation}

\section*{Acknowledgments}
We gratefully acknowledge funding from the European Research Council (ERC) under the European Union’s Horizon 2020 research and innovation programme through an ERC starting grant (DJK, No. 758383). 

\section*{Author contributions}
J.M., M.v.H. and D.J.K. designed the research and experiments; J.M. conducted the experiments; J.M. and S.H. performed simulations and theoretical modeling; all authors analyzed the data, discussed the results, and wrote the paper. 

\section*{Competing interests}
The authors declare no competing interests.

\section*{Additional Information}
Correspondence and requests for materials should be addressed to D.J. Kraft.

\end{document}


\section{Supplementary Figures}

\begin{figure*}[h]
    \centering
    \includegraphics[width=\textwidth]{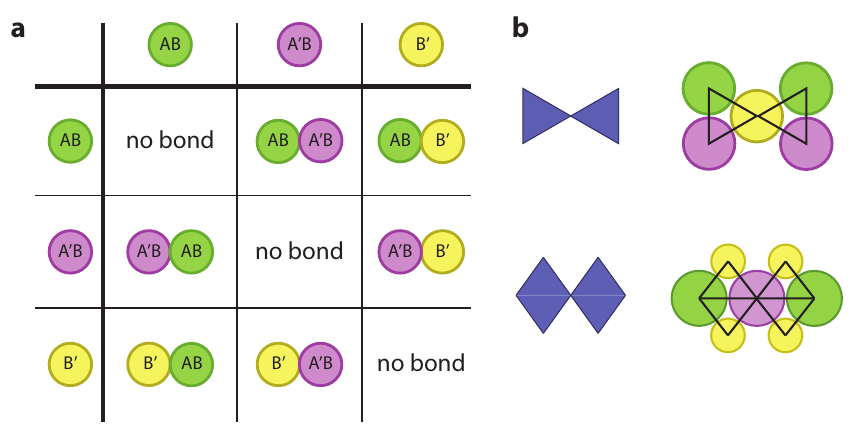}
    \caption{\textbf{DNA functionalization scheme.} a) Interaction matrix of particles functionalized with double stranded DNA strands A, A', B, and B' as indicated. Particles have been functionalized with linkers AB (green), A'B (purple), and B' (yellow) such that they can form bonds between their complementary strands, indicated by primes. b) This interaction matrix makes it possible to create mechanically stable triangular and diamond shaped units in colloidal pivots using the depicted design.}
    \label{fig:ext_dat_fig1}
\end{figure*}

\begin{figure}
    \centering
    \includegraphics[width=0.75\textwidth]{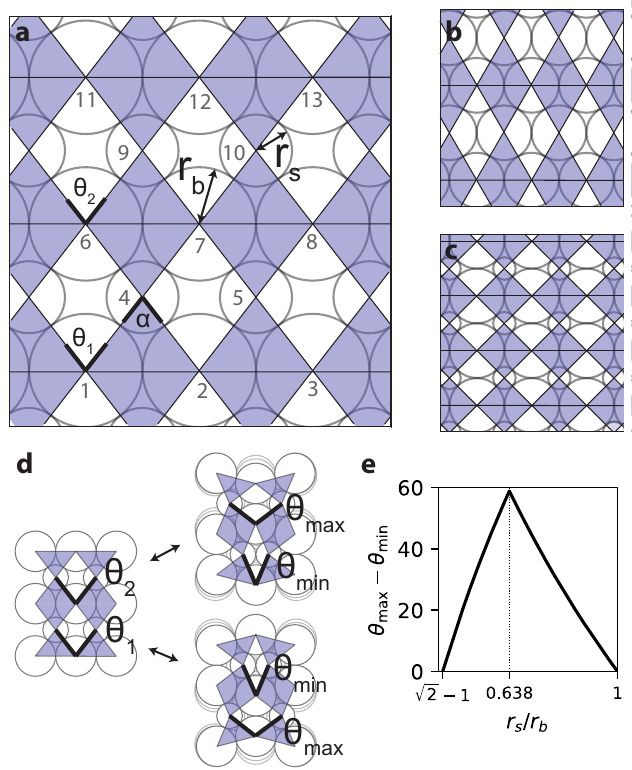}
    \caption{\textbf{Design principles of rotating diamond structures.} (a) Schematics of rotating diamond lattices with particles of size $r_s$ and $r_b$ with angles $\theta_1$, $\theta_2$, and $\alpha$ indicated. (b) Monodisperse particles (i.e. for $r_s/r_b=1$) lead to hexagonally closed packed structures and rigidity. (c) The structure also rigidifies for the lower size limit, i.e. for $r_s=(\sqrt{2}-1)r_b$. (d)  Since a rotating diamond structure is a mechanism, the angles $\theta_1$ and $\theta_2$ are anticorrelated, i.e. when $\theta_1=\theta_\text{min}$, $\theta_2=\theta_\text{max}$ and vice versa. (e) The angular range of motion $\theta_\text{max}-\theta_\text{min}$ that is allowed without particle overlap has an optimum at $r_s=0.638r_b$. }
    \label{fig:ext_dat_fig2}
\end{figure}

\begin{figure*}
    \centering
    \includegraphics[width=\textwidth]{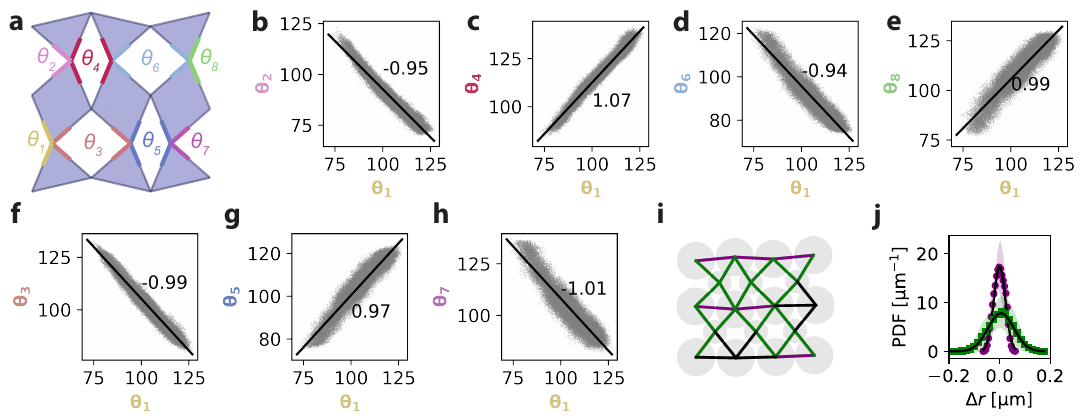}
    \caption{\textbf{Angle correlations rotating diamond structure.} Experimentally measured angle correlations of  a rotating diamond structure with $N=18$ between the colored angles indicated in (a) are plotted in (b-h). For $\theta_3$, $\theta_4$, $\theta_5$ and $\theta_6$, the average of the two opposing indicated angles is calculated and used. The solid black lines represents linear fits with $\theta_i=a\theta_1+b$ with $a$ the slope, whose value is noted in each panel and close to 1 or -1. (i) Particle bonds used to calculate the variation in bond length $\Delta r$ with respect to the mean value $\langle r\rangle$, for which the mean probability density distributions are plotted in (j). The bonds indicated in black in (i) were not used due to a small but noticeable tracking error. Bonds between larger / smaller particles are shown in purple / green, respectively. The shaded region in (j) represents the standard deviation.}
    \label{fig:ext_dat_fig4}
\end{figure*}

\begin{figure*}
    \centering\includegraphics[width=0.95 \textwidth]{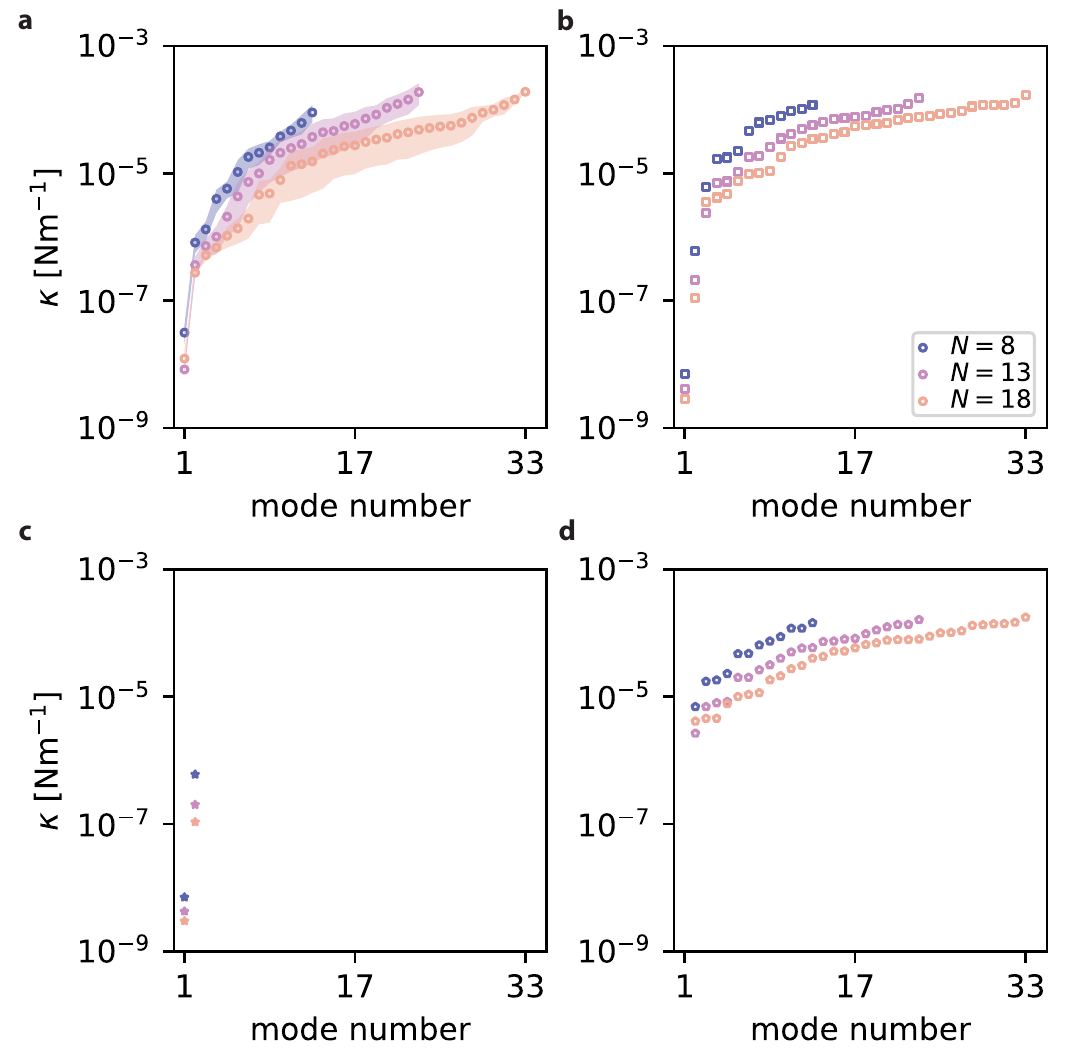}
    \caption{\textbf{Mode spectra for rotating diamond structures.} Mode spectra for (a) experiments, (b) Brownian particle simulations, (c) a Monte Carlo-based conformation sampling approach, and (d) a linear spring model of rotating diamond structures with $N=8,13,18$ as indicated by the color. The mode stiffnesses are the eigenvalues $\kappa_\nu$ obtained from the stiffness matrices (see Methods). Experiments and simulations show that there is one mode significantly softer than the other modes as exemplified by a large jump in stiffness, which is expected from the network structure. Modes obtained from conformation sampling reveal two soft modes that match the softest experimental and simulation modes. Because of a rigid bond assumption, stiff modes cannot be captured. The modes obtained from the linear spring model, which captures extension and compression of bonds, matches well with the stiff modes of the experimental data. The soft modes are strictly zero and hence not visible in the plot. The shaded region in (a) represents the standard deviation.}
    \label{fig:ext_dat_fig3}
\end{figure*}

\begin{figure*}
    \centering
    \includegraphics[width=0.95\textwidth]{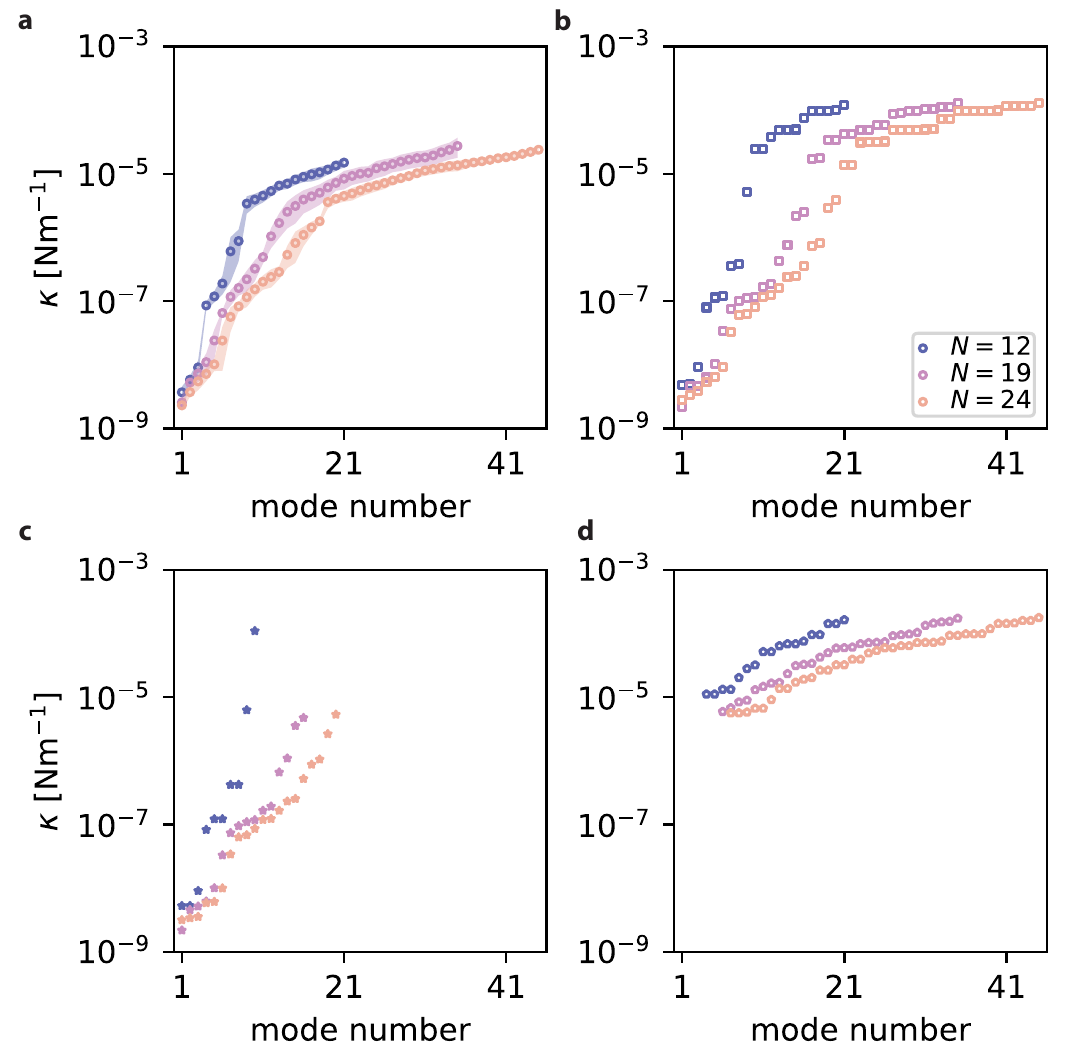}
    \caption{\textbf{Mode spectra for Kagome structures} Mode spectra for (a) experiments, (b) Brownian particle simulations, (c) a Monte Carlo-based conformation sampling approach, and (d) a linear spring model of Kagome structures with $N=12, 19,24$ particles as indicated by the color. In experiments and simulations, 3, 5, and 6 significantly softer modes are found as exemplified by a jump in stiffness, matching with expectations for the number of soft modes based on the bond network.  Note that experimental resolution limits the stiffness values visible in experiment. Modes obtained from sampling all possible conformation describes soft reconfigurations only and shows the nonlinear effects stemming from large soft displacements also present in experiments and simulations. The modes obtained from the linear response model matches the stiffness of the stiffer modes well. The soft modes are strictly zero and hence not visible in the plot. The shaded region in (a) represents the standard deviation.}
    \label{fig:ext_dat_fig5}
\end{figure*}

\begin{figure*}
    \centering
    \includegraphics[width=\linewidth]{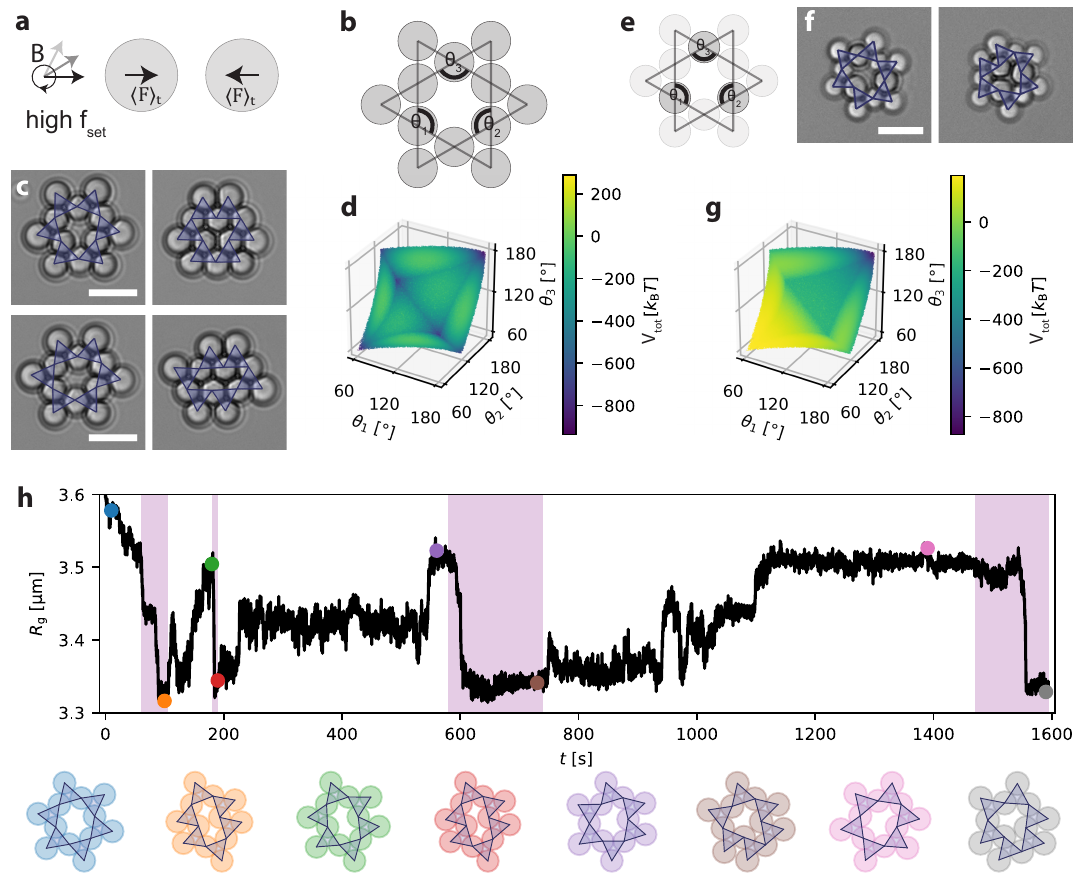}
    \caption{\textbf{Magnetic Kagome structures.} (a) When the rotation  frequency of the magnetic field is higher than the response time of the structure, the particles experience an attractive force $\langle \vec{F}\rangle_t$. (b) Schematic and (c)
    brightfield snapshots of small Kagome structures made from all magnetic particles before (left panels) and after (right panels) applying a rotating magnetic field with 10Hz. The structure folds into two different conformations. (d) The potential energy of calculated as the sum of all dipole interactions shows eight minima when only magnetic particles are used. 
    (e) Schematic and (f) brightfield microscopy snapshots (g) potential energy of a Kagome structure made from three magnetic (darker, slightly bigger) and nine nonmagnetic (lighter, slightly smaller) particles. (f) Upon application of a rotating magnetic field at 10Hz (right panel), the structure folds into a unique conformation. (g) The potential has a single minimum. (h) Radius of gyration of a Kagome structure with all particles being magnetic with conformations at specific times indicated by color. A rotating magnetic field is applied at times shaded in purple, followed by clear drops in the radius of gyration corresponding to a folding of the structure. Scale bars are 5~$\mu$m.}
    \label{fig:ext_dat_fig6}
\end{figure*}

\begin{table*}
\begin{tabular}{ l|l } 
 ssDNA & Sequence \\
 \hline
 ssBB & \makecell[l]{5'-TCG-TAA-GGC-AGG-GCT-CTC-TAG-ACA-GGG-CTC-TCT-GAA-TGT-\\GAC-TGT-GCG-AAG-GTG-ACT-GTG-CGA-AGG-GTA-GCG-ATT-TT-3'} \\
 \hline
 ssA & \makecell[l]{Double Stearyl-HEG-5'-TT-TAT-CGC-TAC-CCT-TCG-CAC-AGT-CAC-CTT-\\CGC-ACA-GTC-ACA-TTC-AGA-GAG-CCC-TGT-CTA-GAG-AGC-CCT-GCC-\\TTA-CGA-\textit{CCT-ACT-TCT-AC}-3'-6FAM} \\
 \hline
 ssA' & \makecell[l]{Double Stearyl-HEG-5'-TT-TAT-CGC-TAC-CCT-TCG-CAC-AGT-CAC-CTT-\\CGC-ACA-GTC-ACA-TTC-AGA-GAG-CCC-TGT-CTA-GAG-AGC-CCT-GCC-\\TTA-CGA-\textit{GTA-GAA-GTA-GG}-3'-Cy3} \\
 \hline
 ssB & \makecell[l]{Double Stearyl-HEG-5'-TT-TAT-CGC-TAC-CCT-TCG-CAC-AGT-CAC-CTT-\\CGC-ACA-GTC-ACA-TTC-AGA-GAG-CCC-TGT-CTA-GAG-AGC-CCT-GCC-\\TTA-CGA-\textit{TAG-TTG-TCA-TT}-3'-6FAM} \\
 \hline
 ssB' & \makecell[l]{Double Stearyl-HEG-5'-TT-TAT-CGC-TAC-CCT-TCG-CAC-AGT-CAC-CTT-\\CGC-ACA-GTC-ACA-TTC-AGA-GAG-CCC-TGT-CTA-GAG-AGC-CCT-GCC-\\TTA-CGA-\textit{AAT-GAC-AAC-TA}-3'-Cy3} \\
 \hline
 ssI & \makecell[l]{Double Stearyl-HEG-5'-TT-TAT-CGC-TAC-CCT-TCG-CAC-AGT-CAA-TCT-\\AGA-GAG-CCC-TGC-CTT-ACG-A-3'} \\
 \hline
 ssI' & \makecell[l]{5'-TCG-TAA-GGC-AGG-GCT-CTC-TAG-ATT-GAC-TGT-GCG-AAG-GGT-\\AGC-GAT-TTT-3'} \\
\end{tabular}
\caption{Overview of single-stranded (ss) DNA strand sequences.  ssBB and either ssA, ssA', ssB, or ssB' are hybridized to form  double-stranded DNA with a single-stranded sticky end (indicated in italic). The sticky ends of ssA and ssA' as well as of ssB and ssB' are complementary and are used for binding colloidal particles. ssI and ssI' are hybridized to form the inert double-stranded DNA strands for steric stabilization. ssDNA strands are modified with dyes 6-Carboxyfluorescein (6FAM) and Cyanine3 (Cy3) at their 3' ends as listed and flexible hexaethylene glycol (HEG) spacers connected to a double stearyl group that inserts into the lipid bilayer thereby providing anchoring for the DNA linkers and inert strands.}
\label{table:dna}
\end{table*}